\documentclass[prc,floatfix,nofootinbib,showpacs,showkeys,reprint]{revtex4-1}

\usepackage[utf8]{inputenc}
\usepackage[T1]{fontenc}
\usepackage{amsmath}
\usepackage{amsfonts}
\usepackage{amssymb}
\usepackage{graphicx}
\usepackage{gensymb}
\usepackage{cleveref}
\usepackage{booktabs}
\usepackage{microtype} 

\usepackage{OwnDefinitions}

\newcommand{\JLU}{Institut f\"ur Theoretische Physik, Universit\"at Giessen, Giessen, Germany}
\newcommand{\HFHF}{Helmholtz Research Academy Hesse for FAIR (HFHF), Campus Giessen, Giessen, Germany}

\begin{document}

\title{Medium modification of pion production in low energy Au+Au collisions}

\author{C. Kummer}
\altaffiliation[Present address: ]{Department of Physics, University of Cyprus, P.O. Box 20537, 1678 Nicosia, Cyprus}
\affiliation{\JLU}

\author{K. Gallmeister}
\affiliation{\JLU}
\affiliation{\HFHF}

\author{L. von Smekal}
\affiliation{\JLU}
\affiliation{\HFHF}

\begin{abstract}
    There is a major mismatch between the charged pion yields in Au+Au collisions at low energies calculated by various transport models and the experimental measured values from the \Hades collaboration.
    In this work, reasonable improvements on the equation of state, in-medium modification of cross sections, and the influence of the nuclear potential for $\Delta$ resonances will be investigated in the framework of the \GiBUU transport model.
    As a result, we demonstrate that theoretical calculations can indeed describe the charged pion yields measured by \Hades for Au$+$Au collisions rather well, but that a mismatch then remains between calculations and data for the yields of neutral pions extracted from dileptons within the same experiment.
\end{abstract}

\maketitle

\section{Introduction}

In the paper by Adamczewski-Musch \etal~\cite{HADES:2020ver} the HADES collaboration showed that transport models systematically overpredict measured pion yields. Every transport code was found  to overshoot the rapidity and $p_t$ spectra by nearly a factor of two. In principle, this problem has been known already for almost 30 years \cite{Bass:1995pj,Teis:1996kx,UmaMaheswari:1997ig,FOPI:1997qpm,Larionov:2001va,Larionov:2003av}. It is disturbing for a few reasons. Due to their low mass, pions play an important role in many theoretical models (see, e.g.~the classic text in Ref.~\cite{Ericson:1988gk}). Assuming two flavors of massless quarks, for example, QCD exhibits a chiral symmetry which, when spontaneously broken, gives rise to three massless Nambu-Goldstone bosons. These are identified with the three pions to explain why they are the lightest hadrons, and hence commonly produced more abundantly than others in heavy-ion collisions. Thus pion production is important from an experimental as well as a theoretical point of view. For the system considered by HADES in Ref.~\cite{HADES:2020ver}, Au+Au at an incident energy of $E_{\rm kin}=1.23 \, A\GeV$, a mismatch between measured and calculated pion multiplicities by not quite a factor of two but on the 50\proz level was reported in Ref.~\cite{Larionov:2020fnu}. Nevertheless, one has to require that transport models reproduce the most frequent particles better than that. It seems therefore very important to develop a framework that can describe the HADES findings and provide an improved  theoretical understanding.

As the conventional assumptions and modes typically employed in transport codes have failed to reproduce the data, new approaches to describe medium modifications of pion production are needed. Indeed, Godbey \etal\ in Ref.~\cite{Godbey:2021tbt} found a prescription to reproduce the correct pion numbers whereby density-dependent suppression factors were applied on the cross sections for each of the charged pions. While this method is phenomenologically certainly successful, the underlying assumption of an exponential suppression with density is currently lacking a sound theoretical basis. Moreover, it seems particularly unclear why an isospin dependent suppression factor should need to be introduced in an otherwise isosymmetric theory.

Here we will therefore compare an isospin-symmetric implementation of this suppression prescription with an alternative approach based on modifications of cross sections through effective in-medium masses \cite{Pandharipande:1992zz,Fuchs:2001fp}. 
%
%
We will first demonstrate that neither of these changes alone are then sufficient to describe the charged pion yields measured by HADES. In combination with a physically motivated modification of the baryonic potential for $\Delta$-resonances, relative to that for nucleons, however, the effective mass suppression of cross sections is able to describe the charged pion yields measured by HADES rather well, and no ad hoc suppression factors are needed.  


Because the suppression of cross sections by effective in-medium masses is sensitive the equation of state (EOS) of nuclear matter, on the other hand, another important prerequisite for a quantitative description of charged pion yields is the right choice of this EOS, of course. Using the relativistic mean-field (RMF) mode of the Giessen Boltzmann-Uehling-Uhlenbeck (GiBUU) transport model, we therefore compare various popular EOSes from the nonlinear (NL) Walecka model. The perhaps most commonly used choice in GiBUU transport simulations of heavy-ion collisions is the set NL2 by Lang \etal\ \cite{Lang:1992jz}, for example. It corresponds to a rather soft EOS, and the resulting effective-mass suppression is not quite strong enough. Interestingly, our results for the charged pion yields from this comparison instead favor parameters by Liu \etal\ \cite{Liu:2001iz} which happen to best satisfy currently available astrophysical constraints, especially those from neutron star mergers \cite{Li:2021thg,Xie:2020tdo}, at the same time.

The structure of this paper is as follows: In \Cref{sec:Model} we review the RMF mode of the GiBUU transport model. Although most of this can be found in the review of Ref.~\cite{Buss:2011mx}, it helps to recall the assumptions and implementations relevant for the modifications implemented in this project. We then first focus on comparing different equations of state against astrophysical constraints and identifying the one by Liu \etal\ \cite{Liu:2001iz} as the presently best choice, before we introduce our in-medium modifications. How these modifications affect particle production and baryon densities is explored in \Cref{sec:Predictions}. Then in \Cref{sec:Comp with experiment} a summary of the results shows how different modifications compare to experiments. These are the FOPI experiment for proton and pion data \cite{FOPI:2004orn,FOPI:2010xrt}, and HADES for pions \cite{HADES:2020ver} and dielectrons \cite{HADES:2019auv}. A summary of our findings is given in \Cref{sec:Conclusions}.

\vspace*{.4cm}

\section{Medium Modifications\label{sec:Model}}

\subsection{Relativistic Mean-Field Theory}

In the present work, the propagation of particles is carried out using a relativistic mean field (RMF) for the equations of motion. Its workings are elaborated here, as it might be less common than non-relativistic Skyrme-like potentials typically used in heavy-ion collisions. More details can be found in the GiBUU review paper \cite{Buss:2011mx}.
The Lagrangian is given by
\begin{widetext}
\begin{equation}
    \mathcal{L} = \overline{\psi} [ \gamma_\mu ( i\partial^\mu -
    g_\omega \omega^\mu - g_\rho \boldsymbol{\tau} \boldsymbol{\rho}^{\,\mu} -
    \frac{e}{2}
    (1+\tau^3) A^\mu )  -  m_N - g_\sigma \sigma ] \psi + \frac{1}{2}\partial_\mu\sigma\partial^\mu\sigma - U(\sigma)
    +  \frac{1}{2} m_\omega^2 \omega^2  + \frac{1}{2}
    m_\rho^2 \boldsymbol\rho^{\,2} - \frac{1}{16\pi} F_{\mu\nu} F^{\mu\nu}~,
\label{eq:Lagr}
\end{equation}
\end{widetext}
including Dirac-spinors $\psi$ for the isodoublet of nucleons with mass $m_N$, a scalar field $\sigma$, as well as isoscalar and isovector vector fields $\omega$ and  $\boldsymbol{\rho}$. 
The self-interactions of the scalar field are parametrized by
\begin{equation}
  U(\sigma) = \frac{1}{2} m_\sigma^2 \sigma^2 +  \frac{1}{3} g_2 \sigma^3
  +  \frac{1}{4} g_3
  \sigma^4 \,,
\end{equation}
with additional coefficients $g_2$ and $g_3$.
The meson-nucleon couplings $g_\omega$, $ g_\rho$, $g_\sigma$ and the mass $m_\sigma$ of the scalar field in the Lagrangian \eqref{eq:Lagr} are determined by the equation of state (EOS) at hand, while the values of $m_\rho$ and $m_\omega $ are conventionally kept fixed at the masses of the physical $\boldsymbol\rho$ and $\omega$ mesons. As usual, $\boldsymbol{\tau}$ are the Pauli matrices for isospin, $A^\mu$ is the electromagnetic field and $F_{\mu\nu}=\partial_\mu A_\nu - \partial_\nu A_\mu$
the corresponding field-strengths tensor.  The Lagrangian \eqref{eq:Lagr} leads to the equations of motion for the Dirac-spinor $\psi$,
\begin{align}
      \Big[ \gamma_\mu \Big( i\partial^\mu - g_\omega \omega^\mu - g_\rho
  \boldsymbol{\tau} \boldsymbol\rho^{\,\mu} -\frac{e}{2} (1+\tau^3)A^\mu \Big)&\nonumber\\
  - m_N -  g_\sigma \sigma \Big] \psi &=0\ ,                          \label{eq:DiracEq}
\end{align}
for the isoscalar scalar field $\sigma$,
\begin{align}
    \partial_\mu\partial^\mu\sigma + \frac{\partial
    U(\sigma)}{\partial\sigma}
  &= - g_\sigma  \rho_S\ ,
  \label{eq:KGsigma}
\intertext{for the isoscalar vector field $\omega$,}
     m_\omega^2 \omega^\nu
  &=\ g_\omega j^\nu_b\ ,
  \label{eq:KGomega}
\intertext{for the isovector vector field $\boldsymbol{\rho}$,}
      m_\rho^2 \boldsymbol{\rho}^\nu
  &=\  g_\rho \boldsymbol{j}_I^\nu\ ,
  \label{eq:KGrho}
\intertext{and for the electromagnetic potential $A$,}
      \partial_\mu\partial^\mu A^\nu  &=\ 4 \pi e
  j_c^\nu\ .
\end{align}
Instead of the explicit Dirac spinors of Eq.~\eqref{eq:DiracEq}, the evaluations in GiBUU are formulated in terms of the corresponding particle distribution functions $f_i(x,\boldsymbol{p})$ which also determine the right-hand sides in the equations of motion above (with kinetic momenta $p^*$ and effective masses $m^*$, as defined below).
These are the scalar density,
\begin{align}
\rho_S
  &=\frac{g}{(2\pi)^3} \sum_{i=p, n, \bar p, \bar n}
  \int\,\frac{ \mathrm{d}^3 p }{ p_i^{*\,0}}\,m_N^*\,f_i(x,\boldsymbol{p})\ ,
\intertext{the baryon density,}
j^\mu_b
  &=\frac{g}{(2\pi)^3}
  \Bigg(\sum_{i=p, n}-\sum_{i=\bar p, \bar n}\Bigg)
  \int\, \frac{ \mathrm{d}^3 p }{ p_i^{*\,0}}\,p_i^{*\mu}\,f_i(x,\boldsymbol{p})
  \,
\intertext{the isospin current,}
     \boldsymbol{j}_I^{\mu}
  &=\frac{g}{(2\pi)^3} \sum_{i=p, n, \bar p, \bar n}
  \int\,\frac{ \mathrm{d}^3 p }{ p_i^{*\,0} }\, p_i^{*\mu}\, \boldsymbol{\tau}\,
  f_i(x,\boldsymbol{p})~,
\intertext{and the charge current,}
     j_c^\mu  &= \frac{1}{2}(j_b^\mu + j_I^{3,\mu})\ .
\end{align}
Here, $g = 2$ accounts for the spin degeneracy. Note that there are no kinetic terms for the vector fields $\omega $ and $\boldsymbol\rho$ in Eqs.~\eqref{eq:KGomega} and \eqref{eq:KGrho} which are thus directly proportional to the baryon and isospin currents. This reflects the fact that these are the corresponding Hubbard fields to represent the short-range 
interactions between nucleons whose strengths only determine the ratios $m_\omega/g_\omega$ and $m_\rho/g_\rho$. As such these auxiliary fields do not necessarily have to be interpreted as the propagating physical vector mesons, and choosing their masses for the values of  $m_\omega$ and $m_\rho$ is only a common convention that then defines the dimensionless couplings $g_\omega$ and $g_\rho$.

Moreover, neglecting isospin-mixed nucleon states, the first two components of the isospin current in Eq.~\eqref{eq:KGrho} are set to zero, i.e.\ 
 $j_I^{1,\nu}=j_I^{2,\nu}=0$, 
and only $\rho^{3,\nu} \propto j_I^{3,\nu}$ needs to be calculated. Finally, the last equation of motion relates 
the electromagnetic current $j_c^\nu$ to the electromagnetic field (in the Lorenz gauge). As convenient kinematic quantities in the definitions of the currents one furthermore introduces the effective nucleon mass 
\begin{align}
m_N^* &= m_N +S
\intertext{and its kinetic four-momentum}
p^* &= p -V \, ,
\end{align}
with scalar and vector fields given by
\begin{align}
    S &= \, g_\sigma \sigma \, ,\\
 V^\nu &= \, g_\omega \omega^\nu + g_\rho
  \tau^3 \rho^{3,\nu} +\frac{e}{2} (1+\tau^3)A^\nu \, .
\end{align}
By assuming plane-wave solutions to the Dirac equation, \cref{eq:DiracEq}, one then obtains the dispersion relation for the nucleons in the form
\begin{equation}
    (p^*)^2\,- \,(m_N^*)^2\, =\,0\,.
    \label{eq:dispersionrelation}
\end{equation}
Although here only given explicitly for the nucleons, this RMF prescription is applied to all baryons. The 
electromagnetic potential is included for all particle species.

\subsection{Equations of State}
\label{sec:EoS}

The nuclear EOSes needed for our purposes relate various thermodynamic observables to the density of symmetric nuclear matter. They are all fixed to yield a binding energy per nucleon of $E/A=-16 \MeV$ at the saturation density $\rho_0 = 0.16 \fm^{-3}$. In other aspects they can and do differ quite substantially, however. The historical reasons for this are that they have originally been devised for different purposes. To study heavy-ion collisions with transport simulations, for example, A.~Lang \etal\ \cite{Lang:1992jz} devised parameter sets specifically suitable for this task, while G.A.~Lalazissis \etal\ \cite{Lalazissis:1996rd} needed a different set to mainly work on nuclear structure models, and B.~Liu \etal\ \cite{Liu:2001iz} have put the emphasis on investigating the influence of isovector scalar fields with yet another EOS. Hence the EOSes by Lang are typically expected to be more applicable for a dynamic description of heavy-ion collisions, while the Lalazissis EOS is considered to better describe nuclear ground states. The EOS of Liu \etal\ is also used by Godbey \etal\ in Ref.\ \cite{Godbey:2021tbt}.

\begin{figure}[htb]
  \begin{center}
  \hspace*{\fill}
  \includegraphics[width=0.45\textwidth]{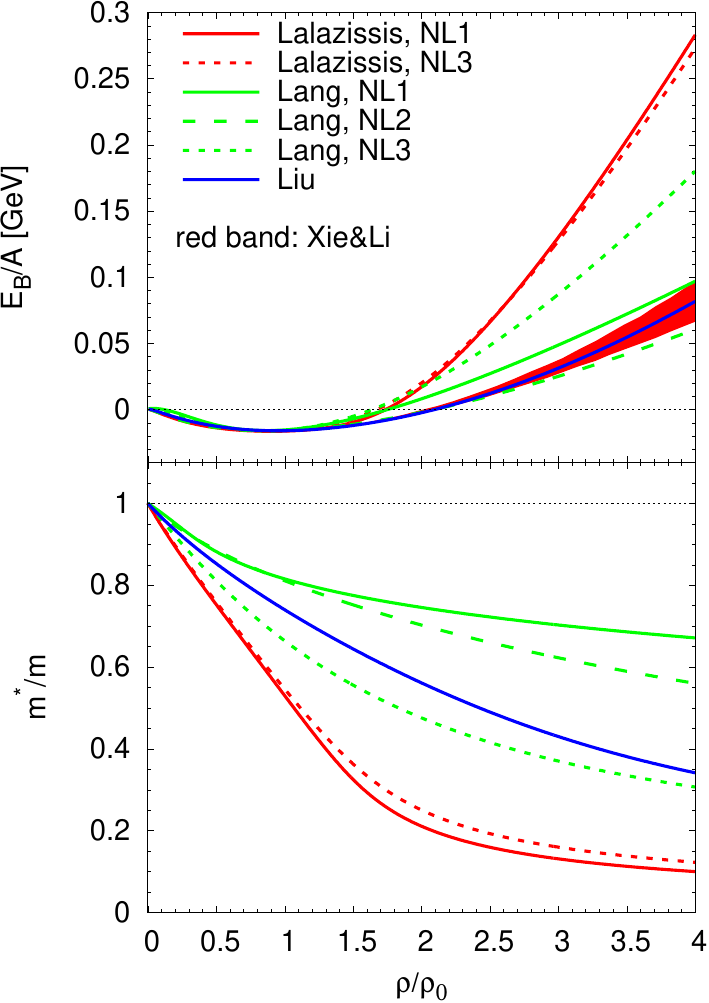}
  \hspace*{\fill}
  \end{center}

  \caption{The binding energy per nucleon and the effective mass as a function of nuclear density for various EOSes \cite{Lalazissis:1996rd,Lang:1992jz,Liu:2001iz}. The red band in the plot of the binding energies shows the reference from astrophysical constraints \cite{Xie:2020tdo}. }
  \label{fig:eoscomp}
\end{figure}

Constraining the nuclear EOS from astrophysical observations, especially on neutron stars has a very long history.
For a rather recent review we exemplarily refer to Li \etal, Ref.~\cite{Li:2021thg}, and the comprehensive list of references given therein. One main emphasis thereby currently is to further constrain the symmetry energy  from current and future neutron-star merger events which provide new data not only on neutron-star masses and radii but also on more complex observables such as their tidal deformability. For our purposes we actually only need the EOS of symmetric nuclear matter which has been studied even more abundantly and is also much more well constrained than the symmetry energy. To be specific, for the comparison of our RMF EOSes we use the reference band of the 68\% confidence interval from astrophysical observations for the EOS of symmetric nuclear matter provided, e.g., in a recent compilation by Xie \etal\  in Ref.~\cite{Xie:2020tdo}. This comparison is shown in the top panel of Fig.~\ref{fig:eoscomp}, where the binding energies per nucleon are plotted together with this reference band over the density in units of the saturation density of symmetric nuclear matter for various EOSes implemented in GiBUU. The bottom panel of Fig.~\ref{fig:eoscomp} shows the corresponding behavior of the density dependent effective masses for the same EOSes in matching colors.
The coefficients from Eq.~\eqref{eq:Lagr} required to reproduce the EOSes by Liu and NL2 by Lang are listed in \Cref{Table:coupl_mass}.

\begin{table*}[tb]
  \centering
  \begin{tabular}{lcccccccc}
    \toprule
    EOS & $K$ & $m_N^*/m_N$ & $g_\sigma$ & $g_\omega$ & $g_\rho$ &
                                                                        $g_2$ & $g_3$ & $m_\sigma$ \\
    & [$\MeV$] & & & & & [$\GeV$] & & [$\MeV$] \\
    \midrule
    Liu & 240 & 0.75 & 8.958 & 9.238 & 3.769 & -4.681 & -30.909 & 550 \\
    Lang, NL2 & 210 & 0.83 & 8.5 & 7.54 & 0.0 & -9.939 & -6.26 & 550.5 \\
    \bottomrule
  \end{tabular}
  \caption{  \label{Table:coupl_mass} Parameter sets for the EOSes by Liu \cite{Liu:2001iz} and NL2 by Lang \cite{Lang:1992jz} together the resulting nuclear incompressibility $K$ and effective nucleon mass $m_N^*$ (in units of their vacuum mass $m_N$) both at saturation.}
\end{table*}

Starting from saturation, where the nuclear incompressibility, cf.~Tab.~\ref{Table:coupl_mass}, which determines the curvature at $\rho_0$, perfectly agrees with the extrapolated value of $K = 240 \pm 20 $~MeV from compression-mode giant resonances in doubly-magic finite nuclei \cite{Garg:2018uam}, we observe that 
the EOS by Liu \etal\ from Ref.~\cite{Liu:2001iz} 
continues to lie right in the middle of the one-sigma confidence interval of the reference curve for all densities up to $4\rho_0$ in Fig.~\ref{fig:eoscomp}. We will therefore mainly consider this EOS by Liu \etal\ from now on. Compared to that, the softer EOS with the NL2 parameters by Lang \etal, commonly used in transport codes, is close to the lower edge of the confidence band for the binding energy. More significantly for our purposes, the fall-off of the effective mass with density is a lot weaker than it is with the Liu EOS, as seen in the ratio $m^*/m$ of the bottom panel in Fig.~\ref{fig:eoscomp}. One therefore expects that the density-dependent suppression of cross-sections by effective masses is less effective with the NL2 Lang parameters than with those for the Liu EOS.  We will show examples of results where this makes a noticeable difference below.

\subsection{Cross-Section Modifications}
\label{sec:CSM}

As mentioned above, transport theoretical calculations overestimate the experimentally measured pion yields in heavy-ion collisions. There are multiple ways to change this. The two main strategies of medium-modifying cross sections that we compare in this work are described next.

\subsubsection{Exponential Suppression\label{sec:expSuppr}}

One way to reduce pion yields is the exponential suppression of cross sections with density in the spirit of Refs.~\cite{Song:2015hua,Godbey:2021tbt}. Here the $NN \to N\Delta$ and $NN \to NN\pi$ cross sections are multiplied by a common factor, i.e.~the same for all isospin channels,
\begin{equation}
    f=\exp\Big(\!-\alpha\,\big({\rho}/{\rho_0}\big)^\beta\, \Big)\ ,
    \label{eq:def expsup}
\end{equation}
with constant $\alpha$ and $\beta$, where the exponent $\beta $ was introduced for more generality. The underlying assumption is that pion distributions can be modeled with this density dependent suppression. In our investigation we have tested various values of $\beta $ and eventually concluded that  $\beta = 1$ can be used to describe the data sufficiently well, with only varying the strength of the exponential suppression with density by adjusting the prefactor $\alpha$. To maintain detailed balance, the factor $f$ is also included in the cross section of the back reaction $N\Delta \to N N$ and in the rate of pion absorption by two nucleons $\pi N N \to N N$.

In this way, the suppression (equally) affects the two dominant channels for pion production. A certain drawback of this method of medium modification on the other hand is that pions are then produced at increased rates by other processes, mainly $NN \to NR$, where $R$ is a higher resonance. While this counteracts the overall reduction of pion numbers to some extend, the  method is nevertheless reasonably effective to describe the measured pion yields.

\subsubsection{Effective Masses\label{sec:eff mass}}

An alternative mechanism for medium modification of cross sections is provided by the suppression through effective masses.

To calculate heavy-ion collisions effectively, one needs to incorporate cross sections for the myriad of different events. One can derive from QFT that the differential cross sections in vacuum are given by
\begin{widetext}
\begin{equation}
    \mathrm{d}\sigma_{12 \to 1'2'\dots N'} =
    (2\pi)^4 \delta^{(4)} \Big(
      p_1+p_2-\sum_{i=1'}^{N'} p_i \Big) \, \frac{n_1n_2\prod_{i=1'}^{N'}
      n_i}{4I_{12}}
\,    \overline{|\mathfrak{M}_{12 \to 1'2'\dots N'}|^2}  \ \mathcal{S}_{1'2'\dots N'} \prod_{i=1'}^{N'} A_i(p_i)
    \frac{\mathrm{d}^4 p_i}{(2\pi)^3 2p_i^{0}}~,
    \label{eq:vacuum cross}
\end{equation}
\end{widetext}
where the symmetry factor
\begin{equation}
    \mathcal{S}_{ab} = \begin{cases}
    1 & \text{if $a$ and $b$ are not identical},\\
    \frac{1}{2} & \text{if $a$ and $b$ are identical} \, ,
    \end{cases}
\end{equation}
takes into account that one cannot distinguish identical particles, and
\begin{equation}
    I_{12} \, = \, \sqrt{(p_1p_2)^2 - (m_1m_2)^2 }
\end{equation} is the flux factor. For Dirac fermions the spin-averaged spectral function is defined as
\begin{equation}
    A(p) =- \frac{1}{g \pi} \text{tr}\Big[\text{Im}\,{S}^{\text{ret}} (p) \gamma^0\Big]
\end{equation} with Dirac matrix $\gamma^0$ and the retarded Green's function ${S}^{\text{ret}}(p)$. Moreover, note that $\mathfrak{M}$ is the matrix element used in the Bjorken-Drell convention, which is related to the PDG  convention $\mathcal M$ by
\begin{equation*}
    \mathcal{M}_{if} = \mathfrak{M}_{if} \prod_j \sqrt{n_j} \ , \quad
    n_j=\begin{cases}
    1 \,  , & \text{$j$ is boson}\, ,\\
    2 m_j\, , & \text{$j$ is fermion}\, .
  \end{cases}
\end{equation*} 
It is important to keep in mind at this point, that not every cross section can be calculated like that from first principles. Many have to be taken from experiment and many arise from educated guesses, when there is not enough data or underlying theory to proceed otherwise.

For the processes relevant to our study, particle number and density are not close to vacuum anymore. Hence it is plausible to introduce some in-medium modifications. There have been many suggestions of how to apply these modifications, the prescription described here is the cross section modification by the effective mass. Following~\cite{Buss:2011mx}, for the in-medium cross section one first writes, 
\begin{widetext}
\begin{equation}
    \mathrm{d}\sigma^*_{12 \to 1'2'\dots N'} =  (2\pi)^4 \delta^{(4)} \Big(
      p_1+p_2-\sum_{i=1'}^{N'} p_i \Big) \, \frac{n_1^*n_2^*\prod_{i=1'}^{N'}
      n_i^*}{4I_{12}^*} \,
    \overline{|\mathfrak{M}_{12 \to 1'2'\dots N'}|^2} \ \mathcal{S}_{1'2'\dots N'} \prod_{i=1'}^{N'} A_i(p_i)
    \frac{\mathrm{d}^4 p_i}{(2\pi)^3 2p_i^{*0}}~,
    \label{eq:inmedium cross}
\end{equation}
\end{widetext} with a corresponding in-medium flux factor
\begin{equation}
    I^*_{12} \, = \, \sqrt{(p^*_1p^*_2)^2 - (m_1^*m^*_2)^2 } \,.
\end{equation}

Evidently, Eq.~\eqref{eq:inmedium cross} is almost identical to Eq.~\eqref{eq:vacuum cross}, except that it contains effective masses and momenta  ($n_j^*$ is analogously defined as $2m_j^*$ for fermions). 
In fact, this is why the Bjorken-Drell convention is used: The matrix element does not need to be in-medium modified, only the other factors are changed. Note however that the delta function has to have the actual four momenta of the particles, as required by energy and momentum conservation.

The trivial suppression of cross sections by the effective masses has been shown to account for the major part of the difference between nucleon-nucleon cross sections in vacuum and in nuclear matter \cite{Pandharipande:1992zz,Fuchs:2001fp}. The simple approximation to keep   $\mathfrak{M}$ constant in the medium is therefore superior to assuming a constant $\mathcal M$ instead. 

The Bjorken-Drell convention of the matrix element on the other hand leads to another problem, as it depends on the so-called free center of mass (c.m.)~energy of the system and thus on the free four-momenta of the colliding particles, while only the kinetic four-momenta are available in the medium. In other words the existing potentials have to be taken into account when calculating the available energy of the system. Thus the c.m.~energy has to be modified as well. One option, which many transport codes (including GiBUU in Skyrme-like mode) use for hadron-hadron reactions, is the free c.m.~energy
\begin{equation}
    s_{\text{free}}\,=\,(p_{1,\text{free}}+p_{2,\text{free}})^2,
\end{equation}
with free momenta
\begin{equation}
    p_{\text{free}}=(\sqrt{m^2 + \boldsymbol{p}^2},\boldsymbol{p})\,.
\end{equation}
This prescription assumes that the potential acts as a background field, which does not affect reaction rates. A drawback of this method is that it cannot account for in-medium thresholds for particle production. One therefore uses a different prescription, namely
\begin{equation}
    \sqrt{s_{\text{free}}}\,=\,\sqrt{s^*}-(m_1^*-m_1)-(m^*_2-m_2)
    \label{eq:sfree}
\end{equation}
with $s^*=(p_1^*+p_2^*)^2$. To show that this prescription conserves in-medium thresholds, one has to start from the assumption, that the sum of vector fields $V$ stays the same before and after the collision, i.e.~that $V_1+V_2 = \sum_{i=1'}^{N'} V_{i}$. One may then replace all four momenta in the delta-function in Eq.~\eqref{eq:inmedium cross} by the corresponding kinetic momenta as well. This is done in the GiBUU code mainly for technical reasons anyway. With this replacement, the cross section therefore becomes proportional to the $N$-body phase-space volume element,
\begin{equation}
\mathrm{d}\sigma^*_{12 \to 1'\dots N'} \propto \mathrm{d} \Phi_N(p_1^*+p_2^*;p_{1'}^*,\dots,p_{N'}^*)\,,
\label{eq:cross phasespace}
\end{equation}
when the outgoing particles are on their in-medium mass shell as defined by Eq.~\eqref{eq:dispersionrelation}. For completeness, the phase-space volume element itself is here defined as usual by
\begin{align}
    \mathrm{d} \Phi_N(P;p_1,\dots,p_N) &= \frac{\mathrm{d}^3 \boldsymbol{p}_1}{(2\pi)^32p_1^0}
    \dots \frac{\mathrm{d}^3 \boldsymbol{p}_N}{(2\pi)^32p_N^0}\label{eq:phasepace_ele_def}\\
    &\quad\times\ \delta^{(4)}(P-p_1-\dots-p_N)\,.\nonumber
\end{align}
Next, one analogously defines the in-medium excess energy $Q^*=\sqrt{s^*}-\sum_{i=1'}^{N'}m_i^*$, and the in-medium threshold condition
\begin{equation}
    Q^*>0
\end{equation}
then follows immediately from Eq.~\eqref{eq:cross phasespace}. Hence it makes sense to first define the in-medium invariant energy as
\begin{equation}
    \sqrt{s_{\text{free}}}=Q^* + \sum_{i=1'}^{N'}m_i = \sqrt{s^*}-\sum_{i=1'}^{N'}(m_i^*-m_i)\,.
    \label{eq:sfree deriv}
\end{equation}
Unfortunately, however, this last equation is hard to compute, as the final particles are not known initially, but produced during the collision. To overcome this problem a second assumption is made, namely that the sum of initial scalar fields also equals the sum of final scalar fields, i.e.~that $S_1+S_2=\sum_{i=1'}^{N'}S_i$, as well. This then finally  allows to rewrite Eq.~\eqref{eq:sfree deriv} into Eq.~\eqref{eq:sfree}. Note that the two assumptions needed in this derivation, namely $V_1+V_2=\sum_{i=1'}^{N'}V_i$ and $S_1+S_2=\sum_{i=1'}^{N'}S_i$, are usually fulfilled in the present simulations, where all baryon-meson couplings are set equal to the nucleon-meson coupling, and the action of the mean-fields on mesons is neglected. Cases when the assumptions do not hold will always be explicitly mentioned and explained below.

All in all it is now possible to calculate the in-medium cross sections for different final states in baryon-baryon collisions. If the final particles are assumed to be on their mass shell, the relation between in-medium and vacuum cross sections is given by
\begin{equation}
    \sigma^*_{12 \to 1'2'\dots N'}(\sqrt{s^*})=\mathcal{F}\,\sigma^{\text{vac}}_{12 \to 1'2'\dots N'}(\sqrt{s_{\text{free}}})
\end{equation}
with the free c.m.~energy given by Eq.~\eqref{eq:sfree}, and  modification factor
\begin{equation}
  \mathcal{F} = \frac{n_1^* n_2^* n_{1'}^* \dots n_{N'}^*}{n_1 n_2
    n_{1'} \dots n_{N'}}
  \frac{I_{12}}{I_{12}^*}
  \frac{\Phi_{N'}(\sqrt{s^*};m_{1'}^*,\dots,m_{N'}^*)}%
  {\Phi_{N'}(\sqrt{s_\text{free}};m_{1'},\dots,m_{N'})}      \,.
        \label{eq:F}
\end{equation}
It contains all factors in Eq.~\eqref{eq:inmedium cross} that depend on  kinetic momenta and effective masses, including the $N$-body phase space volume which is the integrated infinitesimal one in  Eq.~\eqref{eq:phasepace_ele_def},
\begin{equation}
    \Phi_N(M;m_1,...,m_N)= \int \mathrm{d} \Phi_N(P;p_1,...,p_N)\ ,
\end{equation}
with the mass-shell conditions $P^2=M^2$ and $p_i^2=m_i^2$. For baryon-baryon scattering, Eq.~\eqref{eq:F} assumes the form
\begin{equation}
    \mathcal{F} \propto \frac{m_1^*m_2^*}{m_1m_2} \ \prod_{i=1'}^{N'} \frac{m_{i}^*}{m_i}
\end{equation}
which explicitly shows how the ratios of effective to  bare (Dirac) masses lead to the in-medium suppression of cross sections.

Note that not all cross sections need to be modified. As argued previously in the literature, cf.~Refs.~\cite{TerHaar:1987ce,Li:1993rwa,Li:1993ef}, we leave the elastic channels untouched, and only modify inelastic cross sections. Moreover, assuming that the $NN \leftrightarrow N\Delta$ and $NN \leftrightarrow NN \pi$ reactions dominate the inelastic collisions, 
we only modify these cross sections. First of all the procedure is then directly comparable to the prescription with exponential suppression factors described in Sec.~\ref{sec:expSuppr} above. Secondly, for cross sections involving higher resonances it can be assumed that they are governed by short-range interactions (heavy-meson exchanges) for which in-medium effects can be neglected.

\subsubsection{Delta-Potential Modification\label{sec:delta pot}}

Another aspect that possibly needs to be refined is the baryon potential. The free c.m.~energy of \cref{eq:sfree} can be rewritten as
\begin{equation}
    \sqrt{s_{\text{free}}} = \sqrt{s^*} - S_{\text{fin}}\ ,
\end{equation}
where $S_{\text{fin}}$ is the sum of scalar potentials acting on the final particles, and $\sqrt{s^*}$ is the invariant energy of the colliding particles including their vector potentials. 

Phenomenologically, the potential of the $\Delta(1232)$ resonance in nuclei should be on the order of $-30$~MeV \cite{Ericson:1988gk}, see also the comment in 
\cite{Mosel:2020zdw}, and hence by a about a factor of $2/3$ less than that of nucleons,  
\begin{equation}
    U_{\Delta} \simeq -30 \MeV \simeq \frac{2}{3}\, U_N \,.
    \label{eq:pot delta nuc}
\end{equation}
Thus, reducing the scalar and vector potentials of the $\Delta(1232)$ by this factor $2/3$ compared to those of the nucleons, the production of deltas is penalized compared to the default of using the same potentials for $\Delta$ resonances and nucleons. In the spirit of Kosov \etal~\cite{Kosov:1998gp}, these modifications are applied by changing the couplings to the scalar and vector fields. For this purpose we define the ratios
\begin{equation}
    r_S = \frac{g_S'}{g_S} \ \text{ and } \ r_V = \frac{g_V'}{g_V}\, ,
\end{equation}
where $g_S$ and $g_V$ are the couplings of the nucleons to the scalar field $S$ and the vector field $V$, while $g_S'$ and $g_V'$ are the corresponding couplings of the $\Delta $ to $S$ and $V$.

In order to explore the effect of reducing the $\Delta$ potential in the nuclear medium, we will simply set $r_S = r_V = 2/3$ in our comparisons below. While this is most readily implemented in the computations, there is a slight inconsistency with this prescription, however. As mentioned above, the in-medium cross-sections are computed from Eq.~\eqref{eq:inmedium cross} with the Dirac delta function evaluated from the kinetic four-momenta $p^*$ for technical reasons. Strictly speaking, this is only valid when $r_V=1$, however, because momentum conservation is not strictly valid  for  $NN \leftrightarrow N\Delta$ otherwise. Although we do not expect this to have a quantitatively important effect,
the precise extend to which it matters should certainly be assessed at some point.
If necessary, a possible further improvement might then be to only modify the scalar potential, i.e.~to keep $r_V=1$ and obtain $U_\Delta\simeq-30\MeV $ by only adjusting $r_S $ appropriately. We will leave this possibility be explored in the future.

\section{Model Predictions\label{sec:Predictions}}

After our detailed comparisons, as described in Sec.~\ref{sec:EoS}, we have mostly used the EOS by Liu \etal~\cite{Liu:2001iz}  in our simulations. The results labeled as ``default'' in the figures were obtained with vacuum cross sections for comparison.
The two kinds of in-medium modifications are denoted in the figures as follows:
In the results labeled 
as ``$\alpha=2.0$'' we have used the exponential suppression described in Sec.~\ref{sec:expSuppr} with a factor $\alpha=2.0$ in Eq.~\eqref{eq:def expsup}. The in-medium modified  results with the suppression of the cross sections from the effective Dirac masses, as defined in Sec.~\ref{sec:eff mass}, for the processes $NN \leftrightarrow N\Delta$ and $NN \leftrightarrow NN\pi$ are denoted by ``$\sigma^*_{\Delta,\pi}$.'' An additional label ``$r_{S,V}$'' indicates that this was combined  with a modified $\Delta$-coupling to scalar and vector mean-fields, with the default value $r_{S,V}={2}/{3}$, as described in Sec.~\ref{sec:delta pot}. 

In order to demonstrate, qualitatively, how the in-medium modifications affect nuclear densities and collision rates, we have run a few simulations for Au+Au at the HADES incident energy of $1.23 \AGeV$ with zero impact parameter, where we expect the largest effect.  Less central events will have fewer participants and more spectators and thus dilute the differences between the different results.

%
\begin{figure}[htb]
    \begin{center}
    \hspace*{\fill}
    \includegraphics[width=0.4\textwidth]{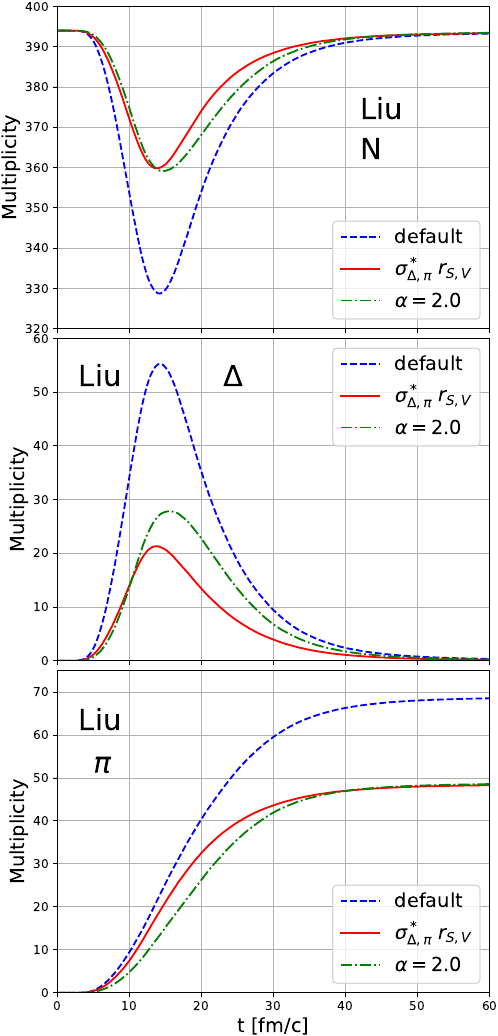}
    \hspace*{\fill}
    \end{center}
    \caption{Particle numbers of nucleons, deltas and pions over time (note the different scales for different particle multiplicities on the $y$-axis, and the offset in that for the nucleons).}
    \label{fig:Liu_multiplicities}
  \end{figure}
The total numbers of nucleons, deltas and pions over time are shown in Fig.~\ref{fig:Liu_multiplicities}. The number of nucleons first decreases through collisions, as other hadrons are being produced by inelastic processes, before back reactions and particle decays finally lead to an increase of nucleon numbers again. Deltas are the resonances of lowest mass and are produced more abundantly than others. Moreover, because they are of particular importance for pion production, they deserve special attention. Since they are not present in the nuclei at the beginning, they have to be created during collisions. Hence, their numbers increase from zero until decays and back reactions destroy almost all the deltas towards the end of the simulation again. The medium modifications are effective in reducing their production and consequently the $\Delta$ yield. The nucleon numbers reflect this behavior as well: When fewer of the most abundant resonances are being produced, more nucleons survive and the dip in the top panel of Fig.~\ref{fig:Liu_multiplicities} is less pronounced. However, the difference gets somewhat mitigated by the enhanced production of higher resonances. Over the relevant time scales during  the heavy-ion collision, pions can be assumed to be stable particles, whose numbers can only be reduced by inelastic collisions. Their production vastly outweighs such processes, however, so it is no surprise that pion numbers only increase over time, here. Just as for the deltas, the in-medium modifications reduce their yields as expected. Note, however, that the different modifications have different effects on the timing of delta and pion formation: Compared to the modification via effective masses and $\Delta$-potential, the exponential suppression has its $\Delta $-multiplicity peak delayed by about $1.6\fm/c$. This is due to the pion capture $\pi N \to \Delta$, which reduces pion numbers in favor of  deltas. The modification of the $\Delta$-potential coupling makes it less likely for deltas to form, so that pion capture is slightly suppressed. In the end, deltas decay into pions, $\Delta \to \pi N$, and the exponential suppression gives a $2.6\fm/c$ delay for pion numbers as compared to the cross-section modification via effective masses and $\Delta$-potential.

\begin{figure}[htb]
    \begin{center}
    \includegraphics[width=.45\textwidth]{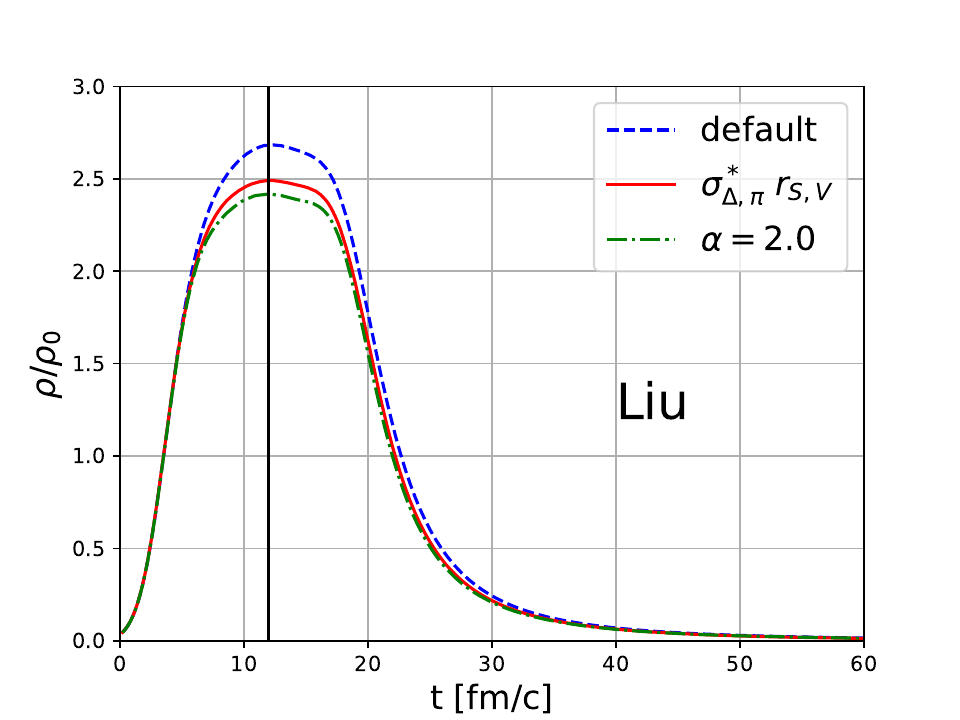}
    \end{center}
    \caption{Central baryon density in units of the nuclear density $\rho_0 = 0.16 \fm^{-3}$ throughout the collision. The vertical line at $t=12 \fm/c$ corresponds to the moment of overlap of both nuclei, thus the moment of maximum density.}
    \label{fig:cbd}
  \end{figure}

The central baryon density is calculated in the center of mass of the system. At the start of the simulations, the nuclei are well separated and the central density is far below the nuclear saturation density of $\rho_0=0.16 \fm^{-3}$. As the two Au nuclei collide, the density increases rapidly up to a maximum at $t=12\fm/c$, when the centers of mass from both nuclei coincide. After that the density rapidly decreases again, until it approaches zero asymptotically. The evolution of the central baryon density over time during this process is shown in Fig.~\ref{fig:cbd}.
Note that the baryon density is not symmetric around its maximum, because stopping slows particles down which causes the expansion to be slower than the compression. Both types of medium modifications lower the central density significantly, e.g.~from $2.82 \, \rho/\rho_0$ without modifications to $2.54 \, \rho / \rho_0$ with the exponential suppression, which amounts to a $10\proz$ decrease. An explanation for this is that less stopping occurs with the reduced cross sections in the medium. Less stopping on the other hand means that the nuclei become more transparent to the nucleons in the reaction zone which are thus more likely to move past one another instead of accumulating in the center region and increasing the density. The effects of the medium modifications on stopping are discussed in more detail in the next section as well.

\begin{figure}[htb]
    \begin{center}
    \hspace*{\fill}
    \includegraphics[width=0.35\textwidth]{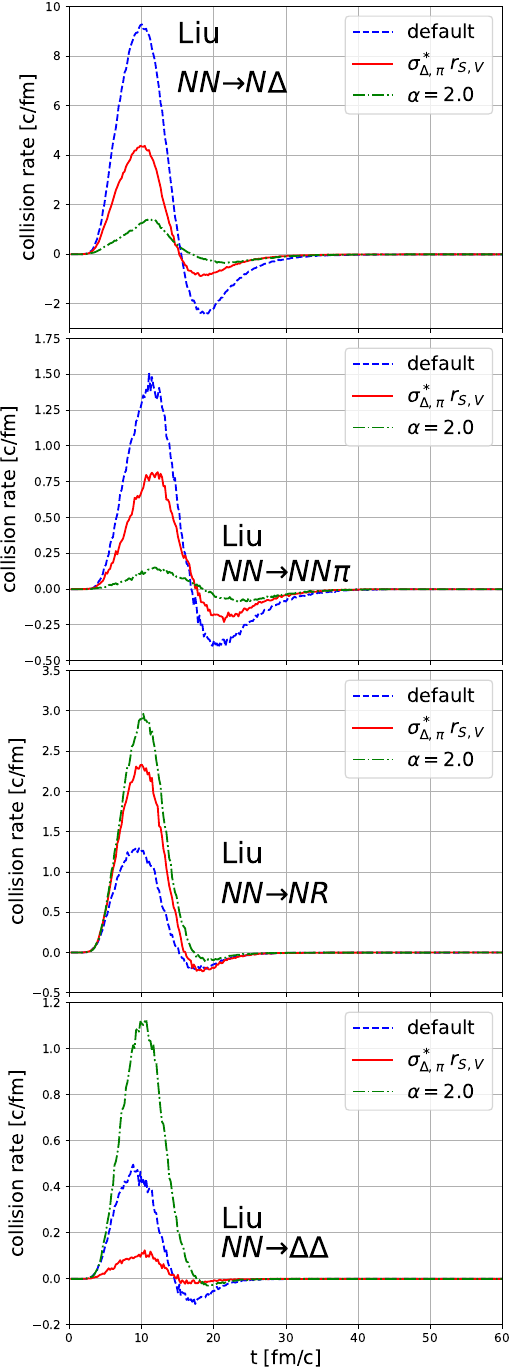}
    \hspace*{\fill}
    \end{center}
    \caption{Comparison of net collision rates for $NN \to N\Delta$, $NN \to NN\pi$, $NN \to NR$, and $NN \to \Delta \Delta$ without (default) and with in-medium modifications.}
    \label{fig:Liu_colrate}
\end{figure}

At the beginning of the collision only nucleons are present, all other hadrons are being produced  by collisions during the reaction. Studying the collision rates can therefore provide an intuitive understanding of the final results. The net collision rates, i.e.~``reaction minus back reaction,'' for the modified processes are compared to the default  rates, using vacuum cross sections, in Fig.~\ref{fig:Liu_colrate}.
Both types of medium modifications significantly reduce the production of $\Delta$ and $\pi$ from the processes $NN\to N\Delta$ and $NN\to NN\pi$.
The exponential suppression factors of Sec.~\ref{sec:expSuppr} in fact almost completely suppress these processes, while the effective masses of Sec.~\ref{sec:eff mass} yield net collision rates somewhere in the middle between the default and the exponentially suppressed ones.  

One might now wonder how total pion numbers can be so similar in both modification scenarios, when the exponential suppression is so much stronger here. The answer lies in the production of higher resonances. As Fig.~\ref{fig:Liu_colrate} also shows, both modifications increase the production of resonances heavier than the $\Delta$, which are collectively labeled $R$ here, in process $NN\to NR$.
This is a direct consequence of our modifications, as more nucleons are present to react and the production of higher resonances becomes more likely if the previously dominant reaction channel is depleted. Because this is more strongly so with the exponential suppression than it is with the  effective masses, see above, the higher resonances are produced even more abundantly with the former as well.

The double delta production, $NN\to\Delta\Delta$, in the bottom panel of Fig.~\ref{fig:Liu_colrate}  is more interesting. As effective as the exponential suppression factors are, in $NN\to N\Delta$ and $NN\to NN\pi$ above, here especially the vastly reduced collision rate in the channel $NN \to N \Delta$, cf.~the top panel in Fig.~\ref{fig:Liu_colrate}, leads to a strong increase in $NN \to \Delta\Delta$, for the same reason as it does in the production of the higher resonances. This is of course unphysical, and due to only reducing the two dominant channels for pion production with the exponential suppression factors. As we can see here, the rate of double delta production in the bottom panel then almost reaches the size of that of a single delta in the top panel of Fig.~\ref{fig:Liu_colrate}. This clearly demonstrates that the strategy of simply suppressing, by explicit factors,  the in-medium cross-sections of the two dominant pion production channels alone needs to be amended, if it is further pursued in the future. 

While the same should be true, in principle, also for the suppression via the in-medium effective masses, here this is overcompensated by the reduction of the $\Delta$ potential in the nuclear medium, cf.~Sec.~\ref{sec:delta pot}. While the effective mass modification alone would enlarge the double delta production rate slightly, although not nearly as much as the exponential suppression does, the reduced $\Delta$-couplings to scalar and vector mean-fields in total clearly lead to a suppression of the double delta production rate as well, as compared to the unmodified default one. The combined effect of effective masses and reduced $\Delta$-couplings therefore yields the overall more realistic phenomenological description in this regard.

\section{Comparison with Experimental Data\label{sec:Comp with experiment}}

We compare our calculations to two experiments, namely HADES \cite{HADES:2020ver} and FOPI \cite{FOPI:2004orn,FOPI:2010xrt}. The pion yields by HADES provided the original motivation for our study, and their dielectrons serve as a good benchmark for our transport model, in addition. The FOPI collaboration measured pions and protons, whereby especially the latter are ideally suited to calibrate the mean-field potentials.

\subsection{Nucleons and Pions from FOPI}

The data for nucleons and pions obtained by the FOPI collaboration are presented in the papers by Reisdorf \etal \cite{FOPI:2004orn,FOPI:2010xrt}. In the experiment \atom{197}{Au} projectiles were collided with stationary \atom{197}{Au} targets with a kinetic energy of 400\MeV and 1.5\GeV per nucleon, respectively.
The FOPI measurements are particularly valuable here, because they include cumulated protons instead of clusters like deuterons or helium-nuclei which GiBUU cannot produce. Hence the predictions from our model are directly comparable to the data.
The FOPI collaboration present their data as functions of longitudinal and transverse rapidities, $y_z$ and $y_x$. The first one corresponds to the usual definition of rapidity along the beam axis, while the second one measures the rapidity in a direction  that is transverse to the beam axis in the laboratory \cite{FOPI:2010xrt},\footnote{The name ``transverse rapidity'' for $y_x$ might potentially be misleading here, because it neither refers to the transverse momentum $p_t$ nor the reaction plane, but solely to a fixed laboratory $x$-direction perpendicular to the beam axis.}
\begin{equation}
    y \equiv y_z = \frac{1}{2} \log \bigg(\frac{E+p_z}{E-p_z}\bigg)
    \, ,\;\;
    y_x = \frac{1}{2} \log \bigg(\frac{E+p_x}{E-p_x}\bigg)\, .
\end{equation}
Moreover, the data published in Ref.~\cite{FOPI:2004orn} is represented in terms of a normalized rapidity $y_{(z,x)}^0\equiv y_{(z,x)}/y_{\text{proj}}$, relative to the projectile's rapidity $y_{\text{proj}}$ in the c.m.~frame.
Finally, only central events were considered, with a scaled impact parameter  $b^0 \equiv b/b_{\text{max}}$ of values up to $b^0=0.15$ which corresponds to impact parameters of up to $b=2.0 \fm$ when $b_{\text{max}}=1.15(A_P+A_T)^{\frac{1}{3}} \fm$ is used as conventional for the maximum impact parameter.

\begin{figure}[b]
    \begin{center}
    \hspace*{\fill}
    \includegraphics[width=0.23\textwidth]{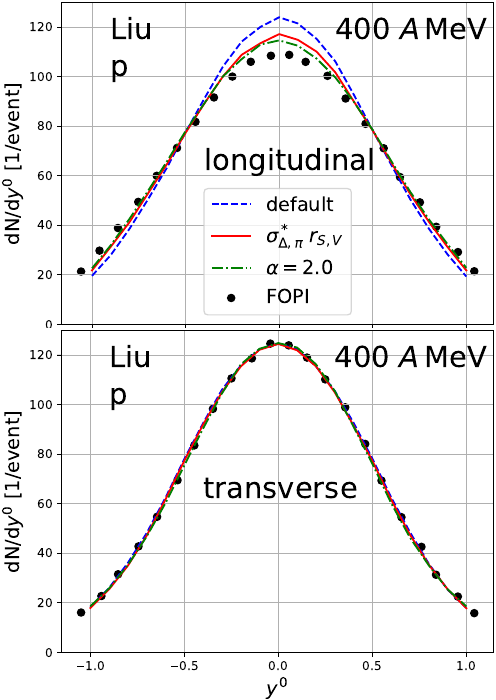}
    \hspace*{\fill}
    \includegraphics[width=0.23\textwidth]{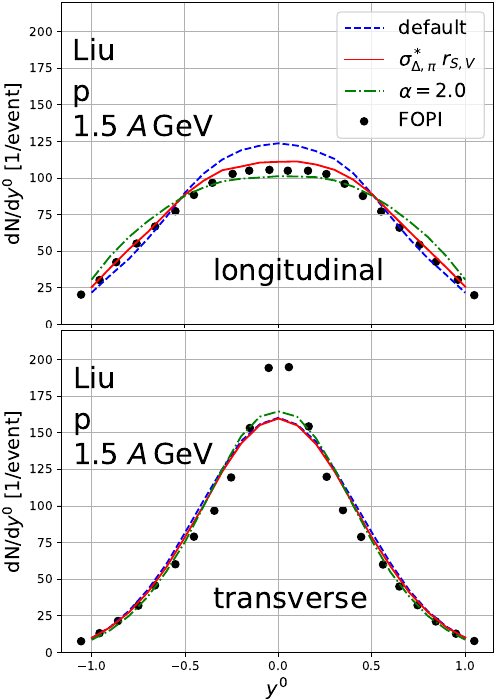}
    \hspace*{\fill}
    \end{center}
    \caption{Proton rapidity spectra without (default) and with in-medium modifications (see text) for beam energies of 0.4\AGeV (left) and 1.5\AGeV (right) compared to the experimental data from Ref.~\cite{FOPI:2004orn}.}
    \label{fig:FOPI_Liu}
\end{figure}

The rapidity spectra of protons for both beam energies are shown in Fig.~\ref{fig:FOPI_Liu}. 
As seen there, our transport calculations (based on the EOS by Liu \etal) describe the experimental data by and large rather well, even with the default vacuum cross-sections.  If anything, the in-medium modifications according to the two different prescriptions both tend to yield slight improvements. The agreement between calculations and data is improved in particular in the longitudinal direction at midrapidity, where the curves are less peaked with the in-medium modifications. This implies that less stopping occurs in the collisions, which is expected when the cross-sections are reduced by the in-medium modifications. 
Comparing the forward/backward regions with the midrapidity region, we furthermore observe that the best overall description is obtained with the effective masses and reduced $\Delta $-couplings, especially at the higher of the two energies, where the finite-density effects in the nuclear medium start to matter in the first place.
The transverse rapidity spectra remain almost unaffected by the in-medium modifications, with minute changes only at the higher energy, which are consistent with the reduced stopping observed in the longitudinal rapidity distribution when the in-medium cross-sections are applied.

\begin{figure}[htb]
    \begin{center} 
    \vspace*{.6cm}
    \hspace*{\fill}
    \includegraphics[width=0.23\textwidth]{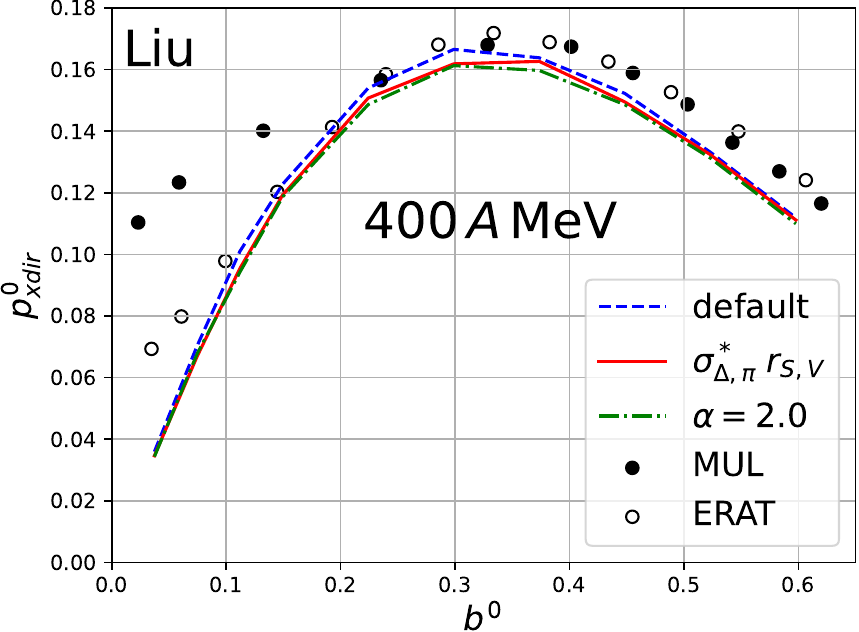}
    \hspace{\fill}
    \includegraphics[width=0.23\textwidth]{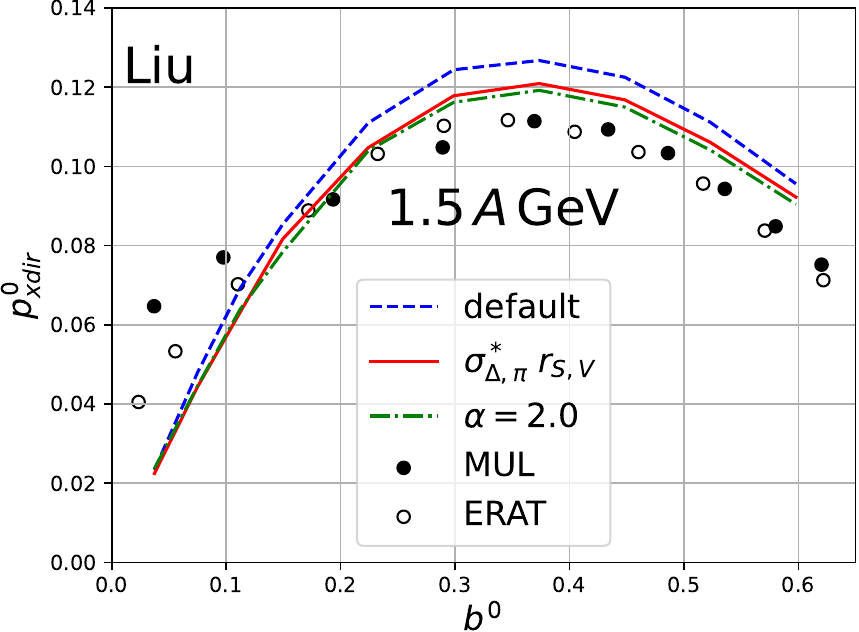}
    \hspace*{\fill}
    \end{center}

    \caption{Sideflow obtained with the EOS by Liu \etal~\cite{Liu:2001iz} with or w/o in-medium modifications for beam energies of 0.4\AGeV (left) and 1.5\AGeV (right) compared to the experimental data from Ref.~\cite{FOPI:2004orn}. MUL and ERAT are the two prescriptions used by FOPI to determine the impact parameter.}
    \label{fig:FOPI_Sideflow_Liu}
%
    \begin{center}
    \vspace*{.6cm}
    \hspace*{\fill}
    \includegraphics[width=0.23\textwidth]{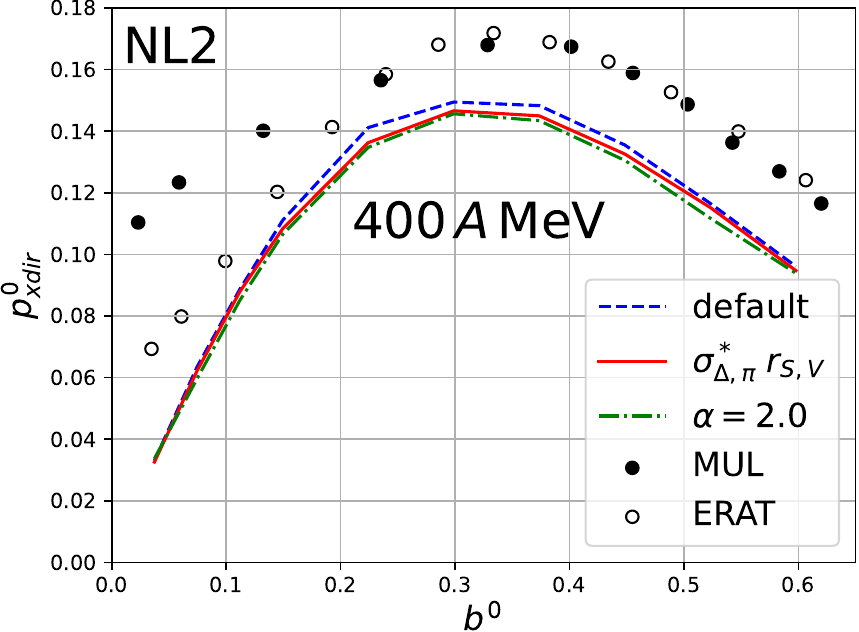}
    \hspace*{\fill}
    \includegraphics[width=0.23\textwidth]{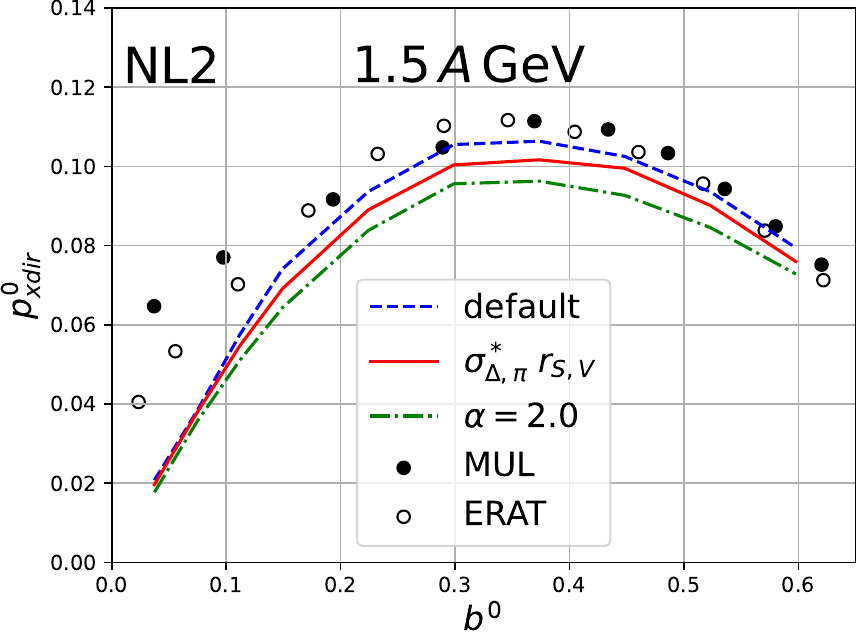}
    \hspace*{\fill}
    \end{center}

    \caption{Same as in Fig.~\ref{fig:FOPI_Sideflow_Liu}, but with NL2 by Lang \etal~\cite{Lang:1992jz} used for the EOS instead.}
    \label{fig:FOPI_Sideflow_NL2}
  \end{figure}

A more complex observable, which is believed to be sensitive to the EOS, is the scaled directed flow, also called sideflow, as introduced by FOPI \cite{FOPI:2004orn} and given by
\begin{equation}
    p_{xdir}^0=\frac{p_{xdir}}{u_{1cm}}\quad,\qquad
    p_{xdir}=\frac{\sum \text{sign}(y)Zu_x}{\sum Z} \ ,
\end{equation}
where the sums run over all charged particles with $Z<10$, excluding pions.
Here $u_{1cm}$ is the spatial part of the center of mass 4-velocity of the projectile. $Z$ the charge of a fragment, $y$ its (longitudinal) rapidity, and $u_x=\beta_x \gamma$ its 4-velocity projected onto the reaction plane. In Figs.~\ref{fig:FOPI_Sideflow_Liu} and \ref{fig:FOPI_Sideflow_NL2} the sideflow is plotted
for our two main EOSes, the one by Liu and NL2 by Lang, 
over the scaled impact parameter and compared to the FOPI data.

The first observation here again is that the in-medium modifications hardly have any effect at all at the lower beam energy of $400\AMeV$, independent of which of the two  EOSes is used in the calculation. The inelastic collisions that are suppressed by the in-medium cross-sections do not matter much, and the influence of the mean fields can therefore be studied more clearly at this lower energy. In particular, we thereby clearly see that the EOS does have an important effect on the sideflow there (independent of which in-medium modifications are applied): While the results from the EOS by Liu \etal~\cite{Liu:2001iz}, with or without in-medium modifications, agree quite well with the data at $400\AMeV$, see the left panel of Fig.~\ref{fig:FOPI_Sideflow_Liu}, those obtained with NL2 by Lang \etal~\cite{Lang:1992jz} significantly underestimate this same data, cf.~the left panel in Fig.~\ref{fig:FOPI_Sideflow_NL2}. 

At the higher energy of $1.5\AGeV$ the calculated sideflow  with the Liu EOS is larger than that with NL2 by Lang, as well. With vacuum cross-sections one would in fact conclude that the latter yields the better description of the data in this case. On the other hand, the in-medium modifications now start to play a more important role in reducing the sideflow at this energy, where more inelastic collisions occur, so that the best overall description of the data, for both energies and including the in-medium effects, is again obtained with the Liu EOS which in all cases produces more sideflow than NL2 by Lang.     

Because the transverse flow should be rather insensitive to
the nuclear incompressibility $K$ \cite{Blaettel:1993uz},  
one would not attribute these differences as predominantly being due to the only slightly larger $K = 240 \MeV $ of the Liu EOS as compared to $K = 210 \MeV $ for NL2 by Lang, cf.~Tab.~\ref{Table:coupl_mass}. In fact, in Ref.~\cite{Nara:2019qfd} it was argued that the effective mass can have a stronger effect than the incompressibility, at least at the higher energy,
because flow is mostly governed by the stiffness at larger densities which is in turn closely related to effective mass and barely dependent on the incompressibility \cite{Boguta:1981px}. Physically, the smaller effective mass of $m_N^* = 0.75\, m_N$ at saturation in the Liu EOS (as compared to $m_N^* = 0.83\,  m_N$ with NL2 by Lang)  requires larger vector repulsion which is here generated by the combined effect of the $\omega $ and  $\boldsymbol{\rho}$ couplings, cf.~Tab.~\ref{Table:coupl_mass}.

\begin{figure}[htb]
    \begin{center}
    \hspace*{\fill}
    \includegraphics[width=0.4\textwidth]{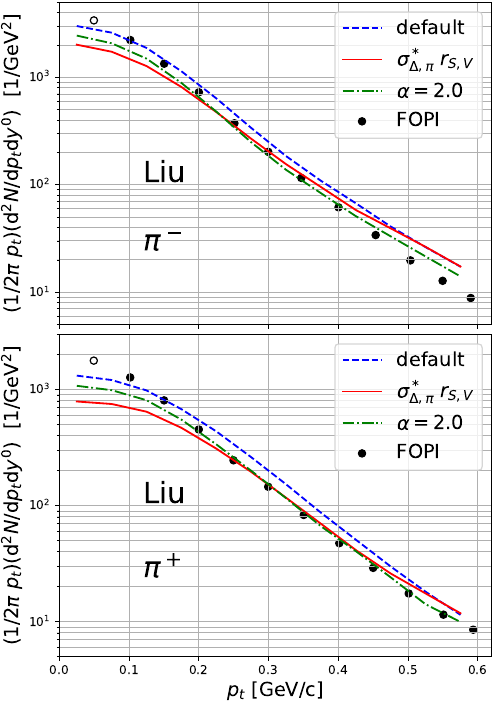}
    \hspace*{\fill}
    \end{center}

    \caption{Transverse momentum spectra of negatively (top) and positively (bottom) charged pions compared to FOPI data at $1.5\AGeV$ \cite{FOPI:2010xrt} for the Liu EOS with and w/o in-medium modifications. The data points shown as empty circles for momenta below  $100 \MeV$ were not measured but extrapolated.}
    \label{fig:FOPI_Pions}
\end{figure}

The FOPI collaboration have also published the transverse momentum spectra for pions \cite{FOPI:2010xrt} at the beam energy of  $1.5 \AGeV$ and for $b^0 \leq 0.15$. Comparisons of our calculations, using the Liu EOS with and without in-medium modifications, with these data are shown in Fig.~\ref{fig:FOPI_Pions}.
The GiBUU calculations using the Liu EOS with vacuum cross-sections agree reasonably well with the lowest measured data points at $p_t\simeq 100\,\text{--}\, 150\MeV$ (note that the points below $100 \MeV $ in the published data have been extra\-polated). While the integrated pion yields (even with the uncertainty of extrapolation at low transverse momenta)  are described correctly by GiBUU in
the default setup, the in-medium modifications lead to some underprediction of the experimental total yields.
Especially the low-momentum pions which dominate the integrated yields are reduced too much to be consistent with the data. In the range of $p_t\simeq 200\,\text{--}\, 400\MeV$ on the other hand, the in-medium modified results can describe the experimental data quite well, while towards higher $p_t$ the slope of the calculated spectra eventually gets too hard as compared to the experimental results.


\subsection{Pions from HADES}

\begin{figure}[htb]
    \begin{center}
    \hspace*{\fill}
    \includegraphics[width=0.4\textwidth]{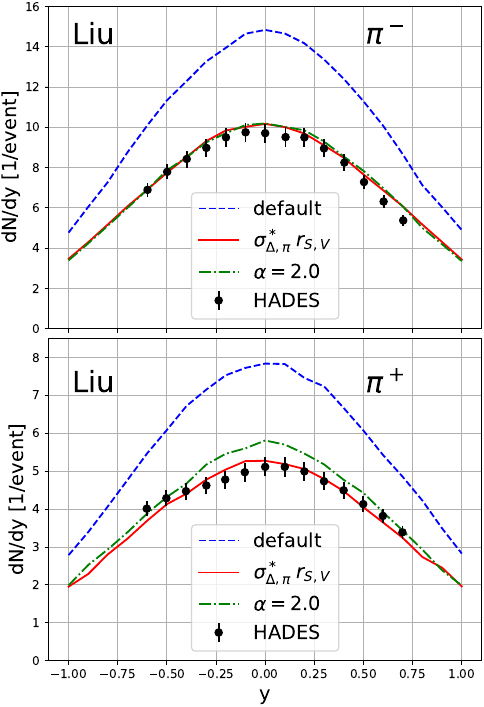}
    \hspace*{\fill}
    \end{center}

    \caption{Rapidity spectra of negatively (top) and positively (bottom) charged pions for the Liu EOS with and w/o in-medium modifications compared to the experimental data from HADES in Ref.~\cite{HADES:2020ver}.}
    \label{fig:HADES_Liu}
\end{figure}

For the HADES experiment, Au+Au heavy-ion collisions were performed at an incident energy of $1.23\AGeV$ and pion spectra were published by Adamczewski-Musch \etal \cite{HADES:2020ver}. Right off the bat, the pion multiplicities are significantly lower than measured by FOPI, which was also discussed by both collaborations \cite{HADES:2020ver,FOPI:2006ifg}. The following comparisons of our GiBUU calculations to the experimental spectra are for the innermost 10\proz centrality class of events \cite{HADES:2017def}. As shown in Fig.~\ref{fig:HADES_Liu}, one observes that both methods of in-medium modifications reduce the pion yields and the corresponding rapidity curves then agree almost perfectly well with data, in particular the one with effective masses and reduced $\Delta$-couplings.

\begin{figure}[h]
    \begin{center}
    \includegraphics[width=.45\textwidth]{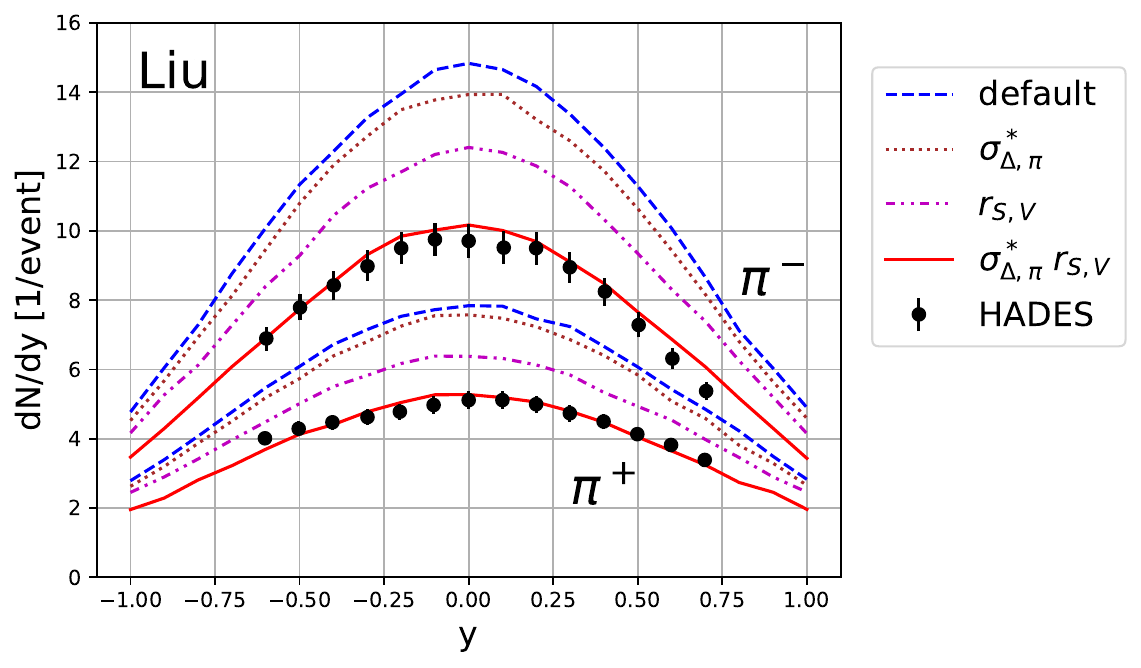}
    \end{center}
    \caption{Rapidity spectra of charged pions, upper(lower) set for $\pi^-$ ($\pi^+$), w/o in-medium modifications (default) compared to separately only either introducing the $m^*/m$-suppression or the modified $\Delta$-potential, together with their combined effect (solid red) as in Fig.~\ref{fig:HADES_Liu}.}
    \label{fig:HADES_Liu_comp}
%
    \begin{center}
    \includegraphics[width=.45\textwidth]{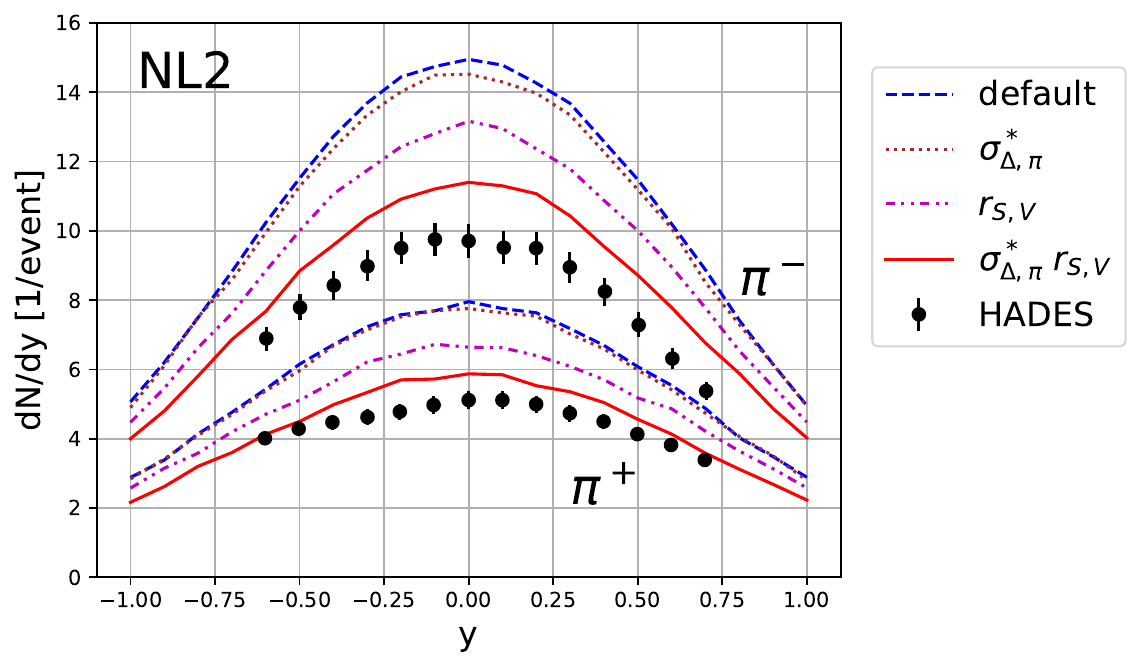}
    \end{center}
    \caption{Same as in Fig.~\ref{fig:HADES_Liu_comp}, but with NL2 by Lang \etal~\cite{Lang:1992jz} used for the EOS instead.}
    \label{fig:HADES_NL2_comp}
\end{figure}

For the latter, it is important to note that it is the interplay between the reduction of the in-medium cross-sections by the effective masses and the modification of the $\Delta$-potential which only together produce the required reduction of the charged pions.
This is illustrated in Fig.~\ref{fig:HADES_Liu_comp}, where the same rapidity spectra are compared to separately either only using the reduced in-medium cross-sections from the effective masses or only the modification of the $\Delta$-potential. Especially the effective masses alone hardly reduce the unmodified default yields, but also the effect of the reduced $\Delta$-couplings alone is by far not enough. The combined effect of both modifications together is in fact considerably stronger than one might expect from the individual ones. 

Another general aspect worth mentioning is that the ratio $\pi^-/\pi^+$ is very robust in all our calculations (with the Coulomb potential in effect for all charged particles). Already the default GiBUU calculation without any of the in-medium modifications yields a $\pi^-/\pi^+$ ratio very close to the experimentally measured one (although the absolute numbers are off by about 50\%). The in-medium modifications lead to a reduction of both charged pion species, but leave their ratio by and large unaffected.

\begin{figure}[htb]
    \begin{center}
    \hspace*{\fill}
    \includegraphics[width=0.4\textwidth]{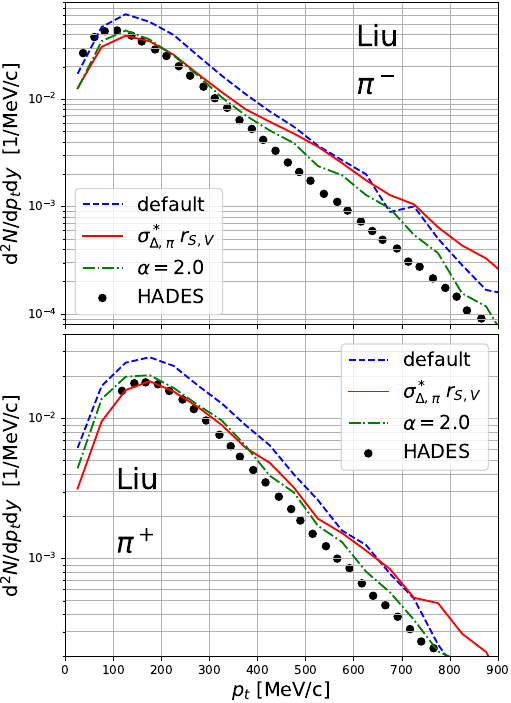}
    \hspace*{\fill}
    \end{center}
    \caption{Transverse momentum spectra of negatively (top) and positively (bottom) charged pions for the Liu EOS with and w/o in-medium modifications (see text) compared to the experimental data from HADES in Ref.~\cite{HADES:2020ver}.}
    \label{fig:HADES_Liu_pt}
\end{figure}

For comparison, the same rapidity curves as in Fig.~\ref{fig:HADES_Liu_comp} but now
with NL2 by Lang \etal~\cite{Lang:1992jz} used for the EOS are shown in Fig.~\ref{fig:HADES_NL2_comp}. First, this shows that the rapidity curves with vacuum cross-sections are almost identical for both EOSes.
The individual modifications by the $m^*/m$ suppression and the reduction of the $\Delta$-potential separately both have similar effects, too.  Remarkably, however, their combined effect is considerably smaller than it is for the EOS by Liu \etal~\cite{Liu:2001iz}. Again, the main difference between the two EOSes is the smaller  $m^*/m$ with the Liu EOS. In fact, the $m^*/m$ effect on the charged pion rapidities with NL2 Lang is even slightly smaller than it is with the Liu EOS. When taken together with the reduction of the $\Delta$-potential, this perhaps unsuspiciously looking difference is apparently enough to produce a sufficient suppression, in a nonlinear fashion, so that only with the Liu EOS the GiBUU calculations with these in-medium modifications perfectly match the charged pion rapidity spectra as measured by HADES.

Similar conclusions are obtained for the exponential suppression with $\alpha=2.0$ of the dominant pion production cross-sections with density as described in Sec.~\ref{sec:expSuppr}: This prescription works for both pion species as well, cf.~Fig.~\ref{fig:HADES_Liu} for the calculation using the Liu EOS, with similar results obtained using  NL2 by Lang. Importantly, with Coulomb potentials included, also the $\pi^-/\pi^+$ ratio again agrees with the experimental data in all calculations, with or without in-medium suppression. This is in contrast to the results of Godbey \etal~\cite{Godbey:2021tbt}, where the factor $\alpha$ needed to be isospin dependent in order to overcome the problem of starting with a wrong ratio already without modifications.

The comparison between calculated and measured transverse-momentum spectra from HADES is shown in Fig.~\ref{fig:HADES_Liu_pt}. As before, the default GiBUU calculation with vacuum cross-sections is compared to the exponential suppression and the combined effective mass plus $\Delta$-potential modification. The $p_t$ spectra calculated with the default GiBUU setup overestimate the experimental data over the entire range of $p_t$ more or less uniformly. Especially the slope of the $\pi^+$ spectrum is in fact described quite well. The transverse-momentum spectrum of the $\pi^-$ falls off too rapidly at very low $p_t$ and is a bit too hard at high $p_t$ in the default calculation. 

One can furthermore see that both types of in-medium modifications explored in this study particularly reduce the pion multiplicities in the lower transverse momentum region. The effects of the in-medium modifications are most visible for transverse momenta $p_t<300\MeV $, while pions with $p_t > 500 \MeV $ remain nearly unaffected. Predominantly reducing low-$p_t$ pions, the calculated spectra get by and large a bit harder by the in-medium modifications, so that 
their slopes in fact reproduce the experimental ones less well than in the default calculation.
Qualitatively, this hardening of the transverse-momentum spectra by the in-medium modifications is in line with our comparison to the pion spectra measured by FOPI, as described in the previous section.

Similarly, for the $\pi^-$ at very low $p_t$, where the spectrum falls off too rapidly in the default calculation here already, the additional in-medium reduction is too strong. The problem that the experimentally measured $\pi^-$ spectra at very low $p_t$ are not well described remains, and if anything, the in-medium modifications make it worse.
An agreement as well as in the case of the rapidity spectra can therefore not quite be obtained between the measured and calculated transverse-momentum spectra of the pions. The differences between the two types of in-medium modifications are not very significant, and this conclusion is not special to us using the Liu EOS here, either.
Overall, however, both types of in-medium modifications 
are effective in reducing the pion yields considerably and  
thus provide a significantly improved description of the experimental data as compared to earlier transport calculations.

\begin{figure}[b]
    \begin{center}

    \vspace{.2cm}
    \hspace*{\fill}
    \includegraphics[width=0.4\textwidth]{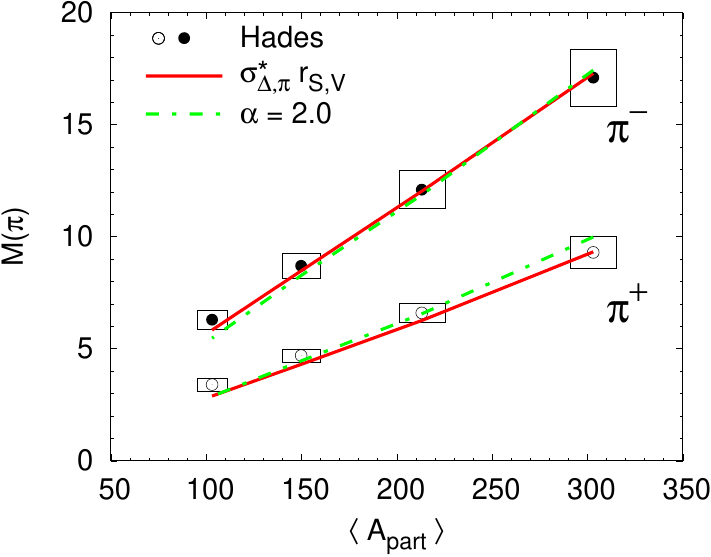}
    \hspace*{\fill}
    \end{center}

    \vspace{-.2cm}
    \caption{Integrated charged pion multiplicities for both types of in-medium modifications (with Liu EOS) over the expected numbers of participants for the various centrality cuts \cite{HADES:2017def} compared to the experimental data from Fig.~7 in Ref.~\cite{HADES:2020ver}. }
    \label{fig:HADES_pimult}
\end{figure}

The detailed comparisons of pion yields as functions of rapidity or transverse momentum described so far were all done for the (0 \text{--} 10\proz) most central bin of the experimental data, where the in-medium effects are expected to be largest. For the integrated yields we have implemented centrality cuts as described in Ref.~\cite{HADES:2017def}.
More specifically, the experimental centrality cuts trigger on the number of participants $A_{\rm part}$, and in this way on the effect of the medium. In line with our original motivation for this analysis, cf.~Fig.~7 in Ref.~\cite{HADES:2020ver}, we have therefore also calculated the total multiplicities of charged pions as functions of $A_{\rm part}$ and compared those to experimental data from HADES in \cite{HADES:2020ver}.
 The results for the two different types of in-medium modifications (both with the Liu EOS) are shown in Fig.~\ref{fig:HADES_pimult}.
One can see clearly that both, the exponential suppression of the 
the in-medium cross-sections for the dominant pion production channels and that by the effective masses plus reduced $\Delta$-potential capture the size dependence of the interaction region extremely well.
There is no way to favor one over the other in terms of the quality of the description of these integrated pion yields. Unlike the transport model calculations included in Fig.~7 of Ref.~\cite{HADES:2020ver}, they both yield total pion multiplicities that are consistent within errors with the experimental data, for both charged pion species and for all centralities.

\begin{figure}[htb]
    \begin{center}

    \vspace{.4cm}
    \hspace*{\fill}
    \includegraphics[width=0.4\textwidth]{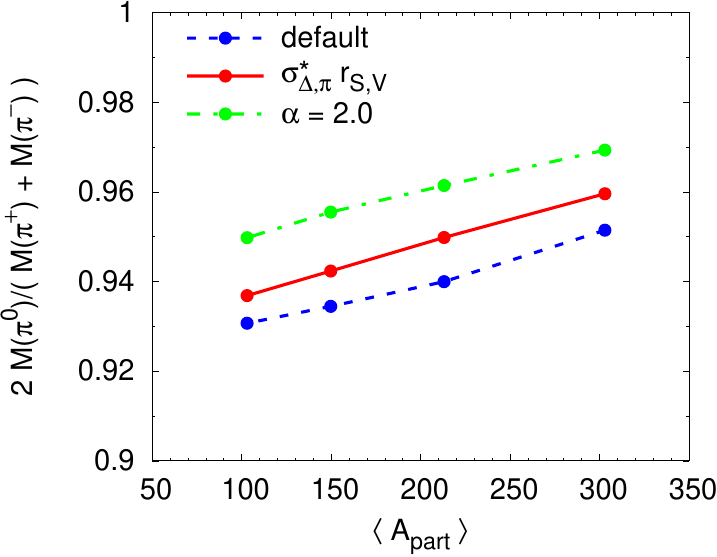}
    \hspace*{\fill}
    \end{center}
    \caption{The ratio of calculated multiplicties of $\pi^0$ over the average $(\pi^++\pi^-)/2$ as function of centrality \cite{HADES:2017def}. }
    \label{fig:HADES_pi0ratio}
\end{figure}

Finally, because the colliding system is strongly charge-asymmetric, the numbers of the different pion charge states can depend on the interacting system size and thus on the impact parameter. While the number of positively and negatively charged pions can in general be very different, the number of neutral pions is often assumed to be well approximated by the average of the two charged pion states. The quality of this assumption is assessed in Fig.~\ref{fig:HADES_pi0ratio}, where the ratio of neutral pions over this average is shown for the different centralities as obtained from our GiBUU calculations with and without the in-medium modifications. 
First, one can see clearly that this ratio always remains close to unity. In particular, our calculations confirm that the average number of positively and negatively charged pions is a valid proxy for that of the neutral ones, at the level of a few percent, and this approximation remains robust with the in-medium modifications included. Moreover, the dependence on the system size is weak and can be neglected as an even smaller effect.

\subsection{Dileptons from HADES}

\begin{figure}[htb]
    \begin{center}
    \hspace*{\fill}
    \includegraphics[width=0.45\textwidth]{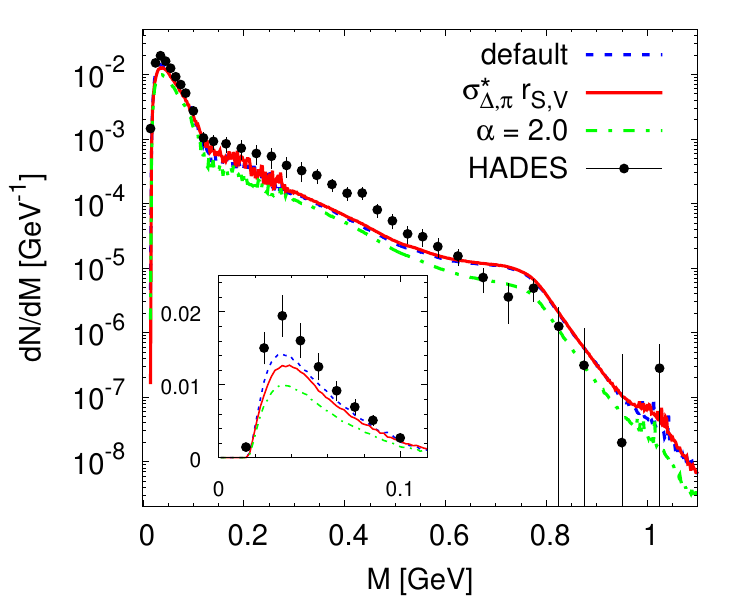}
    \hspace*{\fill}
    \end{center}
    \caption{Invariant-mass spectra of dielectrons for the Liu EOS with and without in-medium modifications of the cross-sections compared to data from \cite{HADES:2019auv}.
    No in-medium broadening of the $\rho $ resonance or $pn$  bremsstrahlungs corrections were included as in \cite{Larionov:2020fnu}, to isolate the effects of an in-medium suppressed pion production here. The insert zooms in on the $\pi^0$-Dalitz region using a linear scale for the plot range.}
    \label{fig:HADES_Dielectrons}
\end{figure}

As the final observable in this study, we compare our GiBUU calculations  with the dielectron spectra measured by HADES.
For Au+Au at $1.23 \AGeV$ these were published by Adamczewski-Musch \etal~in Ref.~\cite{HADES:2019auv}. Our results for the invariant-mass spectrum of dielectrons with the impact parameter distribution corresponding to   $0\, \text{--}\, 40\proz $ centrality are compared to the HADES data in Fig.~\ref{fig:HADES_Dielectrons}. Details of the dilepton calculations with GiBUU can be found in Ref.~\cite{Larionov:2020fnu}. While the main emphasis of this reference was the effect of the collisional broadening of the $\rho$ meson, here we focus on the in-medium suppression of pion production and have therefore deliberately used the vacuum $\rho$-meson spectral function in the calculations.\footnote{The default calculation here corresponds to ``vac.~$\rho$'' in Fig.~20~(a) of Ref.~\cite{Larionov:2020fnu} without the $pn$ bremsstrahlungs correction, and with the Liu EOS instead of NL2 by Lang used there. It was explicitly demonstrated in \cite{Larionov:2020fnu} that the collisional broadening of the $\rho$ together with this correction can fully account for the mismatch between calculation and data in the invariant mass range from roughly $0.2 \GeV$ up to the $\rho$ mass.}

As seen in Fig.~\ref{fig:HADES_Dielectrons}, the effects of the in-medium modifications of pion production on the full invariant-mass spectrum are overall by and large very minor. While there are slight differences in the relative contributions of the various components, the sum of all these components shown here remains nearly unaffected by the in-medium modifications, especially when using the effective masses plus $\Delta$-potential reduction to suppress pion production.  
\REM{Some mystery is the behaviour of the $\pi N$-bremsstrahlung. The in-medium modifications increase the yield of this channel by more than a factor 3. More research is needed to find the source of this enhancement.}

Most interesting for our purposes here is the region around the $\pi^0$-Dalitz peak which is highlighted in the insert of Fig.~\ref{fig:HADES_Dielectrons}.
With the in-medium modifications applied in this study, we observe a suppression of the $\pi^0$-Dalitz decay component in this region. This is not unexpected. As long as the $\pi^0$ yields agree with the average of the charged pions, as shown in the previous subsection, suppressing one without the other is not possible. This is worrisome, however: 
In the best available previous transport calculations of these dielectron spectra in Ref.~\cite{Larionov:2020fnu} the invariant-mass spectrum in this region, where it is dominated by the $\pi^0$-Dalitz decay component, was in fact described quantitatively well, although the charged-pion yields were overestimated by about 50 \%. Therefore, when the in-medium modifications applied here are effective in reducing the total pion numbers, without affecting the ratio of neutral to charged ones, we must inevitably get too few dielectrons from the $\pi^0$-Dalitz decay to describe this region of the experimental data. It seems that solving the first problem, reconciling the calculated charged-pion yields with the experimentally measured rapidity and to some extent also their transverse-momentum spectra, as shown in the last subsection, reveals a new problem. There is now something missing  in the $\pi^0$-Dalitz region of the dielecton rates. For now it seems impossible to get the correct number of dielectrons in the Dalitz region and the correct charged pion yields at the same time.

\section{Conclusions\label{sec:Conclusions}}

As shown in Ref.~\cite{HADES:2020ver}, previous transport-model calculations have so far drastically overpredicted the experimentally measured pion yields. The goal of this paper was to solve this pion puzzle by adjusting the underlying physics of the collisions to in-medium conditions.

Using the relativistic mean-field (RMF) mode of the Giessen
Boltzmann-Uehling-Uhlenbeck (GiBUU) transport model in this report we have first reconsidered the available equations of state in light of recent compilations of astrophysical constraints which allowed us to identify the EOS by Liu \etal~\cite{Liu:2001iz} as the almost ideal choice that
presently best satisfies these phenomenological constraints. With this essentially settled, we have then introduced three different methods for the in-medium modification of pion production.

The first and perhaps most obvious one is to 
simply multiply the cross-sections of the dominant pion production channels $NN \leftrightarrow N\Delta$ and $NN \leftrightarrow NN\pi$ by density-dependent exponential suppression factors in the spirit of an earlier study by Godbey \etal \cite{Godbey:2021tbt}. This works for us as well, with the important difference that in contrast to Ref.~\cite{Godbey:2021tbt} we do not need these factors to be isospin dependent. The likely reason is that the
experimental ratio of the charged pions is reproduced in our default GiBUU calculations already, without any of the in-medium modifications studied here. Therefore, our in-medium suppression of pion production can very well be isospin independent. 

A more physically motivated in-medium suppression of the pion-production cross-sections is provided by the effective Dirac masses in the medium which have previously been shown to account for the major difference between
nucleon-nucleon cross-sections in vacuum and in nuclear matter \cite{Pandharipande:1992zz,Fuchs:2001fp}. Finally, as the third modification, we have reduced the couplings of the $\Delta$ to the scalar and vector mean-fields by a common factor of 2/3 as a simple realization of the reduced strength of the $\Delta$-potential relative to that of the nucleons in nuclei \cite{Ericson:1988gk,Mosel:2020zdw}. Except for a demonstration of their individual effects, we have mostly used the effective masses together with the reduced $\Delta$-couplings at the same time.

We have compared our calculations to experimental data from FOPI \cite{FOPI:2004orn,FOPI:2010xrt} and  HADES \cite{HADES:2020ver,HADES:2019auv}. As a first check, we have made sure that GiBUU can accurately describe the spectra of cumulated protons published by the FOPI collaboration \cite{FOPI:2004orn}. 
These are particularly valuable for our purposes 
because GiBUU does not produce clusters, but only protons. 
One first observation from the proton rapidity spectra is that a soft EOS is needed. 
This is most evident at lower energies ($400 \AMeV$) where the effects of the mean fields can be studied most cleanly, and the stiff EOSes from our comparison can immediately be ruled out. As an example where this sensitivity to the EOS is seen very clearly is the sideflow which favors the EOS
by Liu \etal~\cite{Liu:2001iz} that, as mentioned above, best satisfies the astrophysical constraints at the same time.

That both types of in-medium modifications of pion-production do what they have been designed to do (with the Liu EOS) is seen most clearly in the integrated pion yields where our calculations agree with the data of Ref.~\cite{HADES:2020ver} for both charged pions and all centralities, cf.~Fig.~\ref{fig:HADES_pimult}.

The transverse momentum spectra of pions, as measured by FOPI and HADES, naturally also reflect the reduction of pion yields due to our in-medium modifications. Although the FOPI yields are significantly higher than those from  HADES, which was also discussed by the HADES collaboration in Ref.~\cite{HADES:2020ver}, 
they both show, however, that the in-medium modifications in our calculations tend to reduce low $p_t$ pions more than those at high $p_t$ which remain almost unaffected. As a result, the in-medium modifications studied here generally lead to hardening of the transverse-momentum spectra of the pions which is not supported by the data and which should be addressed in the future.   

Their rapidity spectra as obtained by HADES, on the other hand, are described very well. In particular, with the exponential suppression we do not need different factors for positively and negatively charged pions, as mentioned already. Our comparisons here slightly favor the second type of suppression of pion production, however, from effective in-medium Dirac masses and reduced $\Delta$-potential. Using these together with the Liu EOS we obtain a nearly perfect agreement with the experimental data.
It is thereby important to modify the cross-sections for $NN \leftrightarrow N\Delta$ and $NN \leftrightarrow NN\pi$ and to reduce the $\Delta$-couplings to scalar and vector mean fields. We have demonstrated that the combined effect of the two is surprisingly much stronger than that of the individual modifications. To combine the two modifications is also important for the $\pi^-/\pi^+$ ratio. These two modifications together  
furthermore provide an intuitive theoretical understanding of the  underlying physical mechanisms. And again, the nearly perfect description of the pion rapidity spectra singles out the 
EOS by Liu \etal\ as best satisfying the astrophysical constraints and best describing the HADES data at the same time.

As by Murphy's law the main cause of problems is solutions, 
solving the pion puzzle by our in-medium modifications to pion production creates a new problem here as well:
Comparing our transport calculations with the dielectron spectra measured by HADES \cite{HADES:2019auv} we now observe that 
the strength in the invariant-mass region of 
the $\pi^0$-Dalitz decay goes down as well. At the moment, we are left with the conundrum  that the dilepton yields in the $\pi^0$-Dalitz region are underestimated, if the multiplicites of the charged pions match the experiment, while the traditionally overestimated pion yields, without in-medium modifications, lead to a sufficient strength in the $\pi^0$-Dalitz region to describe the HADES dilepton data.

\begin{acknowledgments}
We are grateful for close collaboration with Alexei Larionov in the early stages of this project. We also gratefully acknowledge helpful discussions with Ulrich Mosel and Jan-Hendrik Otto. We thank Manuel Lorenz, Tetyana Galatyuk and Joachim Stroth for useful questions and comments. 
This work was supported by the German Federal Ministry of Education and Research (BMBF) through Grant No.~05P21RGFCA. 

\end{acknowledgments}

\bibliography{references}

\end{document}